\documentclass{article}

\usepackage{microtype}
\usepackage{graphicx}
\usepackage{subfigure}
\usepackage{booktabs} 

\usepackage{hyperref}

\usepackage[accepted]{synsml2023}

\usepackage{amsmath}
\usepackage{amssymb}
\usepackage{mathtools}
\usepackage{amsthm}

\usepackage[capitalize,noabbrev]{cleveref}
\usepackage[textsize=tiny]{todonotes}

\synsmltitlerunning{Reliable coarse-grained turbulent simulations through combined offline learning and neural emulation}

\begin{document}

\twocolumn[
\synsmltitle{Reliable coarse-grained turbulent simulations through combined offline learning and neural emulation}




\synsmlsetsymbol{equal}{*}

\begin{synsmlauthorlist}
\synsmlauthor{Christian Pedersen}{courant,cds}
\synsmlauthor{Laure Zanna}{courant,cds}
\synsmlauthor{Joan Bruna}{courant,cds}
\synsmlauthor{Pavel Perezhogin}{courant}
\end{synsmlauthorlist}

\synsmlaffiliation{courant}{Courant Institute of Mathematical Sciences, New York University}
\synsmlaffiliation{cds}{Center for Data Science, New York University}

\synsmlcorrespondingauthor{Christian Pedersen}{c.pedersen@nyu.edu}

\synsmlkeywords{Machine Learning}

\vskip 0.3in
]



\printAffiliationsAndNotice{\synsmlEqualContribution} 

\begin{abstract}
Integration of machine learning (ML) models of unresolved dynamics into numerical simulations of fluid dynamics has been demonstrated to improve the accuracy of coarse resolution simulations. However, when trained in a purely offline mode, integrating ML models into the numerical scheme can lead to instabilities. In the context of a 2D, quasi-geostrophic turbulent system, we demonstrate that including an additional network in the loss function, which emulates the state of the system into the future, produces offline-trained ML models that capture important subgrid processes, with improved stability properties.
\end{abstract}

\section{Introduction}

The modelling of turbulent flows is a ubiquitous challenge across many areas of science and engineering. Numerical simulations are always limited to some extent by computational cost. This is of particular concern in the case of turbulent fluids, in which there is significant interaction between length scales. Therefore coarse, but computationally affordable simulations, are often missing important processes that occur below the gridscale.

A widely used approach to tackle this problem is the Large Eddy Simulation (LES) methodology \citep{Lesieur2005,Yang2015}. In this framework, the partial differential equation (PDE) describing the system is smoothed and coarse-grained to a computationally affordable resolution. An additional term referred to as a \textit{subgrid forcing} is added to the coarse resolution PDE which intends to represent the influence of the missing, subgrid dynamics on the resolved field. Models that provide this subgrid forcing term as a function of the resolved field are known as \textit{parameterizations}. 

We focus on the LES formulation in the context of modelling turbulent ocean eddies (\textit{mesoscale} eddies) that play a vital role in climate dynamics \citep{Ferrari2009}. At current computational limitations, these eddies are only partially resolved in coupled climate models \citep{Haarsma2016}, and so including representations of their effects on the resolved field is an essential part of modern climate simulations \citep{FoxKemper2008}. Historically, parameterizations have been constructed via a physical understanding of the unresolved dynamics, such as energy dissipation \citep{Smagorinsky1963,Leith1996} and energy transfer from small to large scales \citep{Jansen2014}. 

Recent years have seen significant work applying machine learning (ML) methods to the problem of turbulence modelling, and more generally solving high-dimensional PDEs. This can take many forms: one setup is to fully learn the system in a purely data-driven approach \citep{Sanchez-Gonzalez2020,Stachenfeld2021}. Alternatively, the classical numerical solver can be augmented \citep{BarSinai2019,Kochkov2021,Sirignano2020} or replaced by ML \citep{Han2018,Sirignano2018}. In the context of the LES framework, ML methods have been used to construct parameterizations turbulent fluids, both in a supervised \citep{Maulik2018,Beck2019} and reinforcement learning setting \citep{Novati2021}. For the modelling of ocean turbulence, parameterizations using convolutional neural networks (CNNs) \citep{Lecun1998} have been trained and successfully incorporated into simulations, forming a hybrid physical + ML model \citep{Bolton2019,Perezoghin2023}.

\begin{figure*}[ht]
\vskip 0.1in
\begin{center}
\centerline{\includegraphics[scale=0.9]{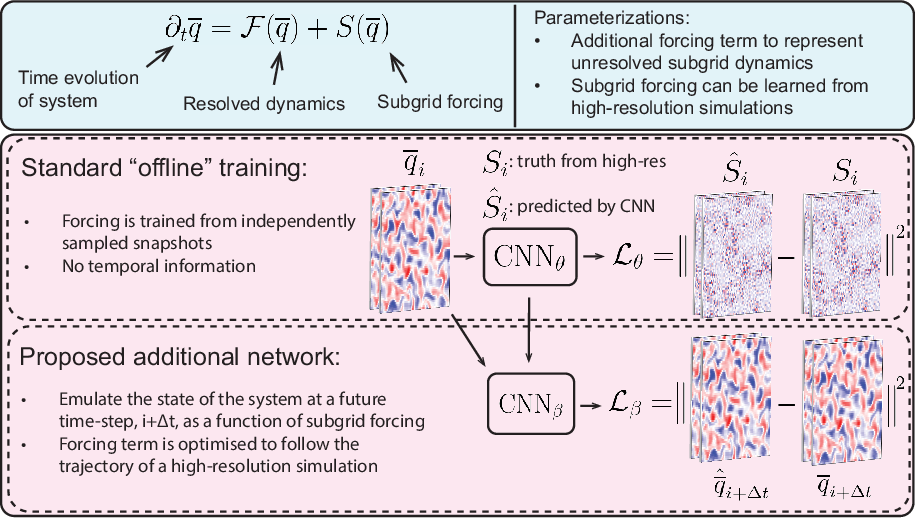}}
    \caption{Overview of the optimisation scheme. A subgrid forcing term, $S_i$, is calculated from a dataset of high-resolution simulations. The standard offline learning framework is to train a CNN to predict this term as a function of the resolved field, $\bar{q}_i$. However this framework contains no information about the temporal evolution of the system. We include an additional network that projects the system forward in time by $\Delta t$. During optimisation of the subgrid model, $\Phi_\theta$, we include this network in the loss function as a regularisation term, and which constrains the system to follow the trajectory taken by a high-resolution simulation.}
    \label{fig:graph}
\end{center}
\vskip -0.1in
\end{figure*}

ML parameterizations can be treated in two categories. In an offline learning setting, a dataset is constructed by running a high-resolution simulation that resolves the target dynamics. These high-resolution fields are smoothed and coarse-grained to obtain low-resolution fields. A subgrid forcing term relating the high- and low-resolution fields can then be calculated (see appendix \ref{app} for details). The ML task is then to model the relationship between a low-resolution field, and the subgrid forcing obtained from high-resolution simulations. However, it has been shown that offline parameterizations often lead to stability issues when implemented in coarse resolution simulations \citep{Maulik2018,Frezat2022,Guan2023}. Some approaches to improving stability of purely offline parameterizations are the inclusion of physics-informed regularisations \citep{Guan2023}, or stochasticity \citep{Guillaumin2021,Perezoghin2023}. In the case of fully-learned systems, stochasticity is also required to produce stable rollouts \citep{Stachenfeld2021}.

An alternative approach is to integrate the ML model within the numerical scheme as a learned correction to each timestep, and train the model over multiple timesteps. This approach is known as online learning, and has been demonstrated to produce better stability properties than purely offline trained models, as well as improved fidelity metrics when compared with high-resolution simulations \citep{Rasp2020,Frezat2022}. This increased performance comes from the fact that the online model is optimised along a trajectory of snapshots in time, rather than learned instantaneously as in the offline setting, where no temporal information is included. However, this procedure requires a differentiable numerical scheme to propagate gradients through. For many applications, including climate modelling, direct integration of the machine learning framework with the numerical scheme is not possible. Therefore, there is motivation to explore approaches that include temporal information in the construction of ML parameterisations of turbulent systems, but in an offline setting (i.e. without requiring interaction with the numerical scheme).

\begin{figure*}[ht]
\vskip 0.1in
\begin{center}
\centerline{\includegraphics[scale=0.5]{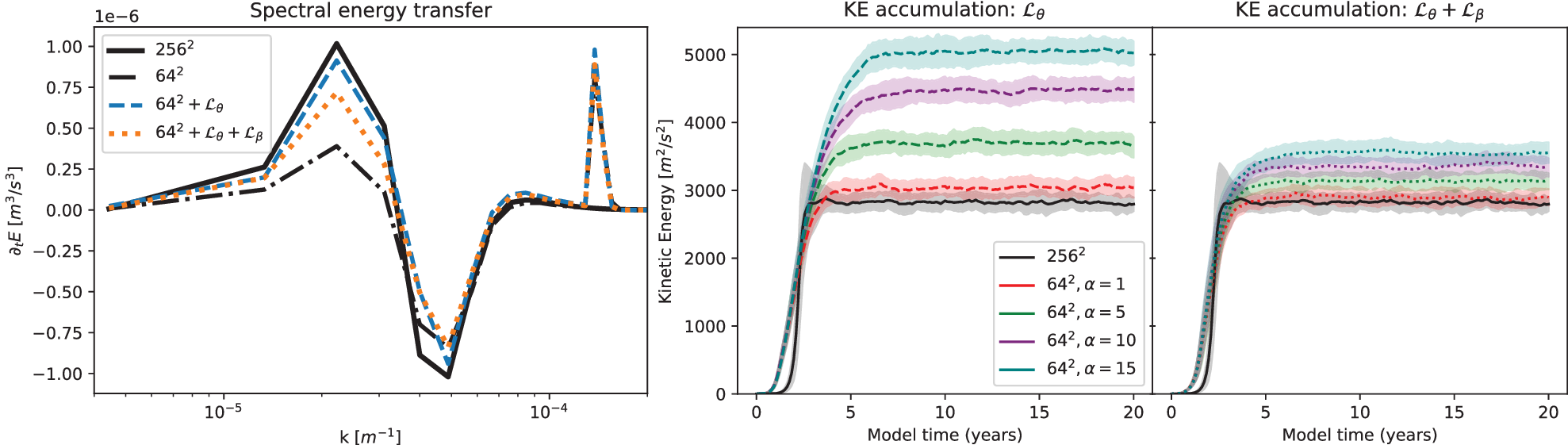}}
    \caption{Online tests comparing the performance of parameterizations constructed in a purely offline mode ($\mathcal{L}_\theta$), and when the additional emulator loss is included ($\mathcal{L}_\theta+\mathcal{L}_\beta$). \textit{Left:} The spectral energy transfer from small to large scales that exists in the high-resolution, $256^2$ simulation, is significantly weaker in the low resolution, unparameterised system ($64^2$). Purely offline parameterizations, capture the positive energy transfer (``backscatter") in the coarse resolution simulation (blue dashed line). Including the emulator loss component (orange dotted lines) significantly improves the backscatter with respect to the unparameterised, coarse system, but not as well as a purely offline loss. \textit{Right:} In order to test the stability of the system, we decrease the diffusivity of the system by increasing $\alpha$, and plot the accumulation of kinetic energy (KE). At $\alpha=1$, the jointly optimised system better reproduces the KE of the high-resolution system. The jointly optimised model is significantly less sensitive to changing $\alpha$, implying improved stability when compared with the pure-offline models.}
    \label{fig:results}
\end{center}
\vskip -0.1in
\end{figure*}

We propose a framework to construct an offline trained model that replicates some of the success of online training. Instead of evolving the system in time using a numerical scheme, we instead add an additional network to the loss function to learn the time dynamics of the system (a process often referred to as \textit{emulation}). By including this network in the loss function and evaluating it on snapshots of high-resolution simulations, we can regularise the ML parameterization during training to follow the trajectory of a high-resolution system. In this way, we have included information about the temporal evolution of the system in the optimisation, without requiring interaction with the numerical scheme \citep{Nonnenmacher2021}. This approach is akin to offline reinforcement learning \citep{Levine2020}, where a policy must be trained from existing datasets, without interacting with the system directly. In such a setting, the learning task includes some learning of the underlying dynamics of the system from the offline dataset. Below we outline the simulation scheme, model architecture, and evaluation metrics.

\section{Methodology}
\subsection{Simulated data}
We use an idealised model of two-dimensional quasi-geostrophic equations (described in Appendix \ref{app}), which assumes a stratified fluid in a rotating system. This system of equations allows for the formation of mesoscale eddies, and has been commonly used to study their parameterizations \citep{Frezat2022,Ross2023}. The prognostic variable of the system is the potential vorticity, $q$, and we denote variables of a coarse-resolution model by $\overline{q}$. We use a system of two fluid layers, implemented in the publicly available code \texttt{pyqg} \citep{Abernathey2022}. The system is advanced using third-order Adams Bashford scheme, and we exploit the fact that the system is in a doubly periodic square domain, by calculating derivatives in Fourier space. This procedure includes a small-scale dissipation filter, which, at each timestep, exponentially damps features below a filter cutoff scale (see appendix \ref{app}). This has the purpose of both removing aliasing noise, and ensuring stability by attenuating the accumulation of energy on small scales, effectively adding diffusion into the system. In order to test stability, we introduce a scaling parameter $\alpha$, which modifies the sharpness of the exponential cutoff in such a way that the diffusivity decreases. We estimate the stability of a given model by the accumulation of kinetic energy in the system, as $\alpha$ is increased (leading to lower diffusivity).

A training dataset can be constructed from an ensemble of high-resolution simulations, outputting a total of $N=18,000$ snapshots. The snapshots are smoothed and downsampled to low-resolution fields, and we build a dataset of tuples: $\{\overline{q}_i,S_i,\overline{q}_{i+\Delta t} \}_{i=1}^{N}$, where $\overline{q}_i$ is the low-resolution, potential vorticity field, $S_i$ is the subgrid forcing calculated from equation $\ref{eq:forcing}$, $\overline{q}_{i+\Delta t}$ is the potential vorticity field at some small future time interval, $\Delta_t$.

\subsection{Model architecture}
Following recent works on ML parameterizations of QG turbulence \citep{Guillaumin2021,Ross2023}, our central ML architecture is a fully-convolutional neural network (FCNN), where we use 5 convolutional layers. We introduce two networks, which contribute to a combined loss function:
\begin{equation}
\mathcal{L}_{\theta,\beta}=\mathcal{L}_\theta+C_\beta\mathcal{L}_\beta;,
    \label{eq:joint_loss}
\end{equation}
where
\begin{equation}
    \mathcal{L}_{\theta}=\frac{1}{n}\sum_i\parallel\Phi_\theta(\bar{q}_i)-S_i\parallel^2
    \label{eq:offline_loss}
\end{equation}
is the standard offline loss term, and
\begin{equation}
    \mathcal{L}_{\beta}=\frac{1}{n}\sum_i\parallel\Phi_\beta({\bar{q_i}},\Phi_\theta(\bar{q_i}))-(\bar{q}_{i+\Delta t}-\bar{q}_i)\parallel^2
    \label{eq:emulator_loss}
\end{equation}
represents the emulator loss. $\Phi_\theta$, a FCNN with parameters $\theta$, predicts the subgrid forcing field $\hat{S}_i$ for a given coarse-resolution field $\overline{q}_i$, and $n$ represents the number of samples in each minibatch. We introduce an additional FCNN, $\Phi_\beta$, which emulates the state of the system at some future timestep $i+\Delta t$, as a function of the both the resolved dynamics $\bar{q}_i$, and subgrid forcing $S_i$. The purpose of this network is to project the parameterization network, $\Phi_\theta$, forward in time, and regularize it in order to have the desired temporal dynamics. This is in contrast to pure offline learning, where the ML task is purely to predict $S_i$ with as high accuracy as possible. Since the dynamics are not closed at the level of the coarse scales, there is an inherent uncertainty in the estimation of the subgrid forcing, irrespective of the capacity of the ML model; in that sense, the role of the emulator is to identify how estimation errors are propagated through time, and use them to steer the system into a stable regime.

We note here that two important design choices had a significant impact on the performance of the parameterization. We found the best performance when jointly optimising the loss function \ref{eq:joint_loss}, but after pre-training the emulator network \ref{eq:emulator_loss}. Secondly, we emulate the residuals of the fields between two timesteps, $(\bar{q}_{i+\Delta t}-\bar{q}_i)$ rather than just $\bar{q}_{i+\Delta t}$. Given that for the small values of $\Delta t$ we are interested in ($\mathcal{O}(10$)), the fields at time $i$ and ${i+\Delta t}$ are extremely similar, and so the majority of the signal in this mapping is simply the identity. We compared using a ResNet architecture to learn this mapping, vs using a FCNN to emulate the residual quantity, and found significantly improved performance using the latter approach. All fields are standardised to have zero mean and unit variance before being passed to the neural networks. When emulating a residual quantity, we found similar performance when re-standardising the residual fields to have unit variance, versus just taking the residuals of the standardised fields. Therefore, for simplicity, we do not re-standardise the residual fields when producing the results in section \ref{sec:results}. We include a scaling coefficient, $C_\beta$, which can be chosen to either ensure the two loss terms are balanced, or put emphasis on either the offline or emulator loss component during training. The $\Phi_\beta$ network predicts the state of the system at some future timestep, $\hat{\overline{q}}_{i+\Delta t}$, where training data is obtained from high-resolution simulations. Therefore this joint optimisation procedure incentivises a subgrid forcing model, $\Phi_\theta(\overline{q})$, which follows the trajectory of a high-resolution simulation.

\section{Results}
\label{sec:results}
The key feature of 2D, quasi-geostrophic turbulence that we aim to capture with parameterizations is the inverse energy cascade \citep{Kraichnan1980}. In the left panel of figure \ref{fig:results}, we show the spectral energy transfer for four different simulations. The solid black lines show results for a high-resolution simulation with $256^2$ grid cells. The positive energy transfer around wavenumber $\kappa=2\times10^{-5}\mathrm{m}^{-1}$, and the negative values around $\kappa=5\times10^{-5}\mathrm{m}^{-1}$ indicate that energy is being removed at small scales, and increased at larger scales following the inverse energy cascade. We refer to the positive energy transfer as a ``backscatter". The low-resolution ($64^2$), unparameterised system in black dash-dot lines, has severely diminished backscatter due to the coarser resolution. This effect is a primary target to be reproduced by a parameterization. In blue dashed lines, we show the spectral energy transfer for a simulation run including a parameterisation trained with a pure-offline loss, i.e. using equation \ref{eq:offline_loss} only. As with previous works \citep{Ross2023}, this baseline model captures the backscatter well.

In orange dashed lines, we show results for a model constructed using the joint loss function, equation \ref{eq:joint_loss}. After experimenting with a range of different time horizons and loss coefficients, we found the best performance in terms of the online metrics shown in figure \ref{fig:results} were achieved using $\Delta t=10$ and $\mathcal{C}_\beta=40$. We note that on average, the values of these two loss functions are different - the $\Phi_\beta$ network is predicting the residual quantity, and as described in the previous section, we do not standarise the residual fields to have unit variance. In addition, the two networks perform differently at their respective learning tasks. We observe that, on average the loss terms $\mathcal{L}_\beta \sim10^{-3}$, and $\mathcal{L}_\theta \sim10^{-1}$. Given this, the $\mathcal{L}_\beta$ term, with a coefficient of $C_\beta=40$, will be approximately an order of magnitude smaller than the $\mathcal{L}_\theta$ term, making the emulator component of the network subdominant, but not irrelevant.

In the right panels, we show the accumulation of kinetic energy over time, as we decrease the diffusivity of the system by increasing $\alpha$. We consistently observed (figure not shown) that this accumulation of kinetic energy occurred around the gridscale, indicating that the parameterization accumulates instabilities. In the rightmost panel, we show results for the jointly optimised model. Firstly, the KE at $\alpha=1$ is more consistent with the high-resolution case. Additionally, as $\alpha$ increases, we see significantly less energy accumulation than with the purely offline model. This indicates that the system with the jointly optimised parameterisation is more stable. We noticed a general trend that the larger the coefficient $\mathcal{C}_\beta$ on the emulator component of equation \ref{eq:joint_loss}, the more significant the gains in stability as indicated by the right panel of figure \ref{fig:results}. However this came at the expense of diminishing the effect of the backscatter. This implies that whilst including temporal information does indeed improve stability, the offline loss term is necessary to capture the desired energy transfer properties. Given this trade-off, we additionally experiment with simply diminishing the contribution of a purely offline trained model. We find that similar results, of diminished backscatter but improved stability, can be obtained by reducing the contribution of a purely-offline model by $20\%$ when tested in online simulations. However this approach would require \textit{a posteriori} tuning of the parameterization, something that our framework does not require.

\section{Summary}
Methods to improve the accuracy of low resolution numerical simulations of turbulent fluids are essential in a wide range of fields, particularly in the case of climate modelling, where important dynamics are only partially resolved. Approaches that require differentiable numerical simulations to include information about the time-evolution of the system (\textit{online} training) are highly effective, but particularly in the case of climate modelling, these methods are incompatible with many key simulators. We have presented a novel ML framework for the parameterization of important subgrid dynamics in ocean simulations, which includes temporal information without requiring differentiable simulations. We demonstrate that this approach leads to improved stability when the ML architecture is included into a hybrid ML+physics simulation, and identify a trade-off between stability properties and capturing the desired energy transfers. There is significant ongoing work to be explored, to clarify questions such as whether pre-training and fixing the emulator weights, or jointly optimising the networks produces optimum performance, and understanding the balance between the time-emulation and offline components of the loss function. Additionally, the emulator network could be used over multiple timesteps, analogous to the methodologies in \cite{Rasp2020} and \cite{Frezat2022}.

\section*{Broader impact}
The primary motivation of this work is to improve representations of subgrid dynamics in the modelling of ocean turbulence. This is a crucial component in modelling of the climate system, and therefore of being able to produce reliable climate projections. The representation of subgrid effects has been identified as one of, if not the major source of discrepancy between different climate projections. Reliable climate modelling is of fundamental importance at the economic, societal and governmental level, as projections inform decisions made in order to prepare for changes to Earth's climate, and so advancements in this area will have significant broader impact.

\bibliography{refs}
\bibliographystyle{synsml2023}

\newpage
\appendix
\onecolumn

\section{Quasi-geostrophic equations solved}
\label{app}
We consider a two-layer, quasi-geostrophic system, with the prognostic variable is the potential vorticity, given by
\begin{equation}
q_m=\nabla^2\psi_m+(-1)^m\frac{f_0^2}{g'H_m}(\psi_1-\psi_2),m\in\{1,2\},
\end{equation}
where $m=1$ denotes the upper layer, $m=2$ denotes the lower layer, $H_m$ is the depth of the layer, $\psi$ is the streamfunction, which is related to the fluid velocity by $\mathbf{u}_m=(u_m,v_m)=(-\partial_y\psi_m,\partial_x\psi_m)$, and $f_0$ is the Coriolis frequency. The time evolution of the system is given by
\begin{equation}
\partial_tq_m+\nabla\cdot(\mathbf{u}_mq_m)+\beta_m\partial_x\psi_m+U_m\partial_xq_m=-\delta_{m,2}r_{ek}\nabla^2\psi_m+\mathrm{ssd}\circ q_m,
\label{eq:q_dt}
\end{equation}
where $U_m$ is the mean flow in the $x$ (zonal) direction, $\beta_m=\beta+(-1)^{m+1}\frac{f_0^2}{g`H_m}(U_1-U_2)$, $r_{ek}$ is the bottom drag coefficient, and $\delta_{m,2}$ is the Kronecker delta. These equations are solved numerically in spectral space, using 3rd order Adams-Bashford integration. The final $\mathrm{ssd}$ term refers to small-scale dissipation:
\begin{equation}
    \mathrm{ssd} =\begin{cases} \text{e}^{-\alpha 23.6\,\left(\kappa^{\star} - \kappa_c\right)^4}: &\qquad \kappa^{\star} \ge\kappa_c\\
\,\,\,\,\,\,\,\,\,\,\,1:&\qquad \text{otherwise}\,.
\end{cases}
\label{eq:ssd}
\end{equation}
where the cutoff is given by $\kappa_c=(0.65\pi)/\Delta x$, and $\kappa^{\star}$ is a radial wavenumber:
\begin{equation}
\kappa^{\star} \equiv \sqrt{ (k\,\Delta x)^2 + (l\,\Delta x)^2 }\, ,
\end{equation}
with $k$ and $l$ being wavenumbers in the horizontal and vertical directions, and $\Delta x$ is the grid size. This dissipation filter is included for the purposes of dealising, and to ensure stability of the system. We include a parameter $\alpha$, by default $\alpha=1$, in order to reduce the diffusivity of the system, and enable tests of stability presented in figure \ref{fig:results}.

We denote a smoothed field as having been convolved with some filter kernel, $G(\mathbf{y})$:
\begin{equation}
\overline{\phi}(\mathbf{x})=\int G(\mathbf{y-x})\phi(\mathbf{y})d\mathbf{y},
\end{equation}
where we use the same filter kernel as in equation \ref{eq:ssd}, with $\alpha=1$. Smoothing equation $\ref{eq:q_dt}$, we get an equation for the low-resolution system:
\begin{equation}
\partial_t\overline{q}_m+\nabla\cdot(\mathbf{\overline{u}}_m\overline{q}_m)+\beta_m\partial_x\overline{\psi}_m+U_m\partial_x\overline{q}_m=-\delta_{m,2}r_{ek}\nabla^2\overline{\psi}_m+S+\mathrm{ssd}\circ\bar{q}_m,
\end{equation}
where now we have an additional forcing term $S$, which accounts for dynamics occurring below the smoothing scale:
\begin{equation}
S=\nabla\cdot({\mathbf{\overline u}}\;\overline{q}-\overline{\mathbf{u}q})\;.
\label{eq:forcing}
\end{equation}

\end{document}